# Modulated magnetic structure of Fe$_3$PO$_7$ as seen by $^{57}$Fe Mössbauer spectroscopy

A.V. Sobolev, A.A. Akulenko, I.S. Glazkova, D.A. Pankratov, I.A. Presniakov

*M.V. Lomonosov Moscow State University, Moscow, Russia*

**Abstract** The paper reports new results of the $^{57}$Fe Mössbauer measurements on Fe$_3$PO$_4$O$_3$ powder sample recorded at various temperatures including the point of magnetic phase transition $T_N \approx$ 163K. The spectra measured above $T_N$ consist of quadrupole doublet with high quadrupole splitting of $\Delta_{300K} \approx 1.10$ mm/s, emphasizing that Fe$^{3+}$ ions are located in crystal positions with a strong electric field gradient (EFG). In order to predict the sign and orientation of the main components of the EFG tensor we calculated the EFG using the DFT approach. In the temperature range $T < T_N$, the experimental spectra were fitted assuming that the electric hyperfine interactions are modulated when the Fe$^{3+}$ spin (***S***) rotates with respect to the EFG axis and emergence of spatial anisotropy of the hyperfine field $\boldsymbol{H}_{hf} \propto \tilde{\boldsymbol{A}} \cdot \boldsymbol{S}$ at $^{57}$Fe nuclei. These data were analyzed to estimate the components of the anisotropic hyperfine coupling tensor ($\tilde{\boldsymbol{A}}$). The large anharmonicity parameter, $m \approx 0.94$, of the spiral spin structure results from easy-axis anisotropy in the plane of the iron spin rotation. The temperature evolution of the hyperfine field $H_{hf}(T)$ was described by Bean-Rodbell model that takes into account that the exchange magnetic interactions are strong function of the lattice spacing. The obtained Mössbauer data are in qualitative agreement with previous neutron diffraction data for a modulated helical magnetic structure in strongly frustrated Fe$_3$PO$_4$O$_3$.

## 1 Introduction

The Fe$_3$PO$_4$O$_3$ (or Fe$_3$PO$_7$) phosphate, known in the literature as the mineral grattarolaite [1], forms a noncentrosymmetric crystal lattice consisting of triangular units (Fe$^{3+}$)$_3$, which are coplanar with the hexagonal (*ab*) planes. The local coordination of Fe$^{3+}$ ions is a distorted trigonal bipyramid (FeO$_5$) cluster. These clusters are arranged in triangular subunits linked to one face within the iron triangle. Below $T_N = 163$ K, previous magnetic and thermodynamic measurements [2] revealed very strong magnetic Fe$^{3+}$ coupling ($\Theta_{CW} \sim -1000$ K) with frustration parameter, $|\Theta_{CW}|/T_N > 6$, indicating significant frustration of the antiferromagnetic interactions. According to the recent powder neutron diffraction data [2], the strong frustration induces an ordered helical incommensurate structure with the helical axis in the hexagonal (*ab*) plane and modulation wave vector of modulus $|\delta| = 0.073$ Å$^{-1}$. It was shown that the wave vector $\boldsymbol{k}_h = (\delta_a, \delta_b, 1.5)$ for the modulation does not change as a function of temperature. There are two types of near-neighbor magnetic exchange interactions (Fig. 1): the nearest-neighbor $J_1$ exchange ($z_1 = 2$) within the triangle (Fe)$_3$, and the $J_2$ exchange ($z_2 = 4$) coupling trigonal units in different *c*-axis layers. According to [2], the observed commensurate antiferromagnetic order along the *c*-axis implies that the $J_2$ exchange is dominant, and the helical modulation within the (*ab*) plane arises as a compromise for the competing $J_1$-$J_2$ interactions.

One of the intriguing features of the magnetic structure in Fe$_3$PO$_4$O$_3$ is the needle-like domains [2]. The small in-plane correlation length (~70 Å) persisting down to the lowest temperatures ($T \ll T_N$) blocks of long-range order of the helical magnetic structure. It was assumed that the appearance of domain walls in the structure is a result of the frustrated $J_1$ interactions within triangular (Fe)$_3$ units [2]. However, the mechanism for stabilization of domains in antiferromagnetic phase is not well understood. Moreover, because of the high concentration of disordered domains, neutron diffraction on the powder does not allow uniquely to determine the orientation of the helix axis within the (*ab*) plane.

In this work, we present the results of the first detailed Mössbauer study of Fe$_3$PO$_4$O$_3$ in a wide temperature range, including magnetic phase transitions. Since previous $^{57}$Fe Mössbauer data were reported only above $T_N$ [3, 4], we performed measurements new experiments down to 15 K in order to complement the study of unusual magnetic structure of this triangle-based material. In the range $T < T_N$, the spectra are analyzed assuming a space-modulated helical magnetic structure proposed in [2]. Such an approach allows us to reproduce, from experimental spectra, the profile of the spatial anisotropy of the hyperfine field, $H_{hf}$ and the large easy-axis anisotropy in the plane of the iron spin rotation. We carried out a detailed analysis of the temperature dependences of hyperfine parameters in light of the peculiarities of the electronic and magnetic states of the iron ions in Fe$_3$PO$_4$O$_3$.

## 2 Experimental

The powder Fe$_3$PO$_4$O$_3$ sample was prepared by two-step solid-state reaction. Firstly, we prepared iron phosphate FePO$_4$ by mixing the stoichiometric amounts of FeC$_2$O$_4$·2H$_2$O and NH$_4$H$_2$PO$_4$, and then two-step annealing was performed at 350°C for 12 h and at 620°C for 12h in air. Secondly, we mixed stoichiometric amounts of Fe$_2$O$_3$ and FePO$_4$ powders and annealed at 950°C for 12h, and finally annealed several times at 1075°C for 12h.

X-ray powder diffraction (XRPD) data were collected at RT on a RIGAKU MiniFlex600 diffractometer using CuK$\alpha$ radiation (2$\theta$ range of 10–80°, a step width of 0.02°, and scan speed of 1 deg/min). The XRPD patterns of the synthesized samples showed the formation of the unique rhombohedral Fe$_3$PO$_4$O$_3$ phase (space group *R3m*). The refined lattice parameters of Fe$_3$PO$_4$O$_3$ in hexagonal reference (*a* = 8.006(1) Å and *c* = 6.863(4) Å) are in good agreement with literature data [3]. In what follows, the rhombohedral Fe$_3$PO$_4$O$_3$ phase will be referred as "Fe$_3$PO$_7$".

Mössbauer experiments were performed in transmission geometry with a 1500 MBq $\gamma$-source of $^{57}$Co(Rh) mounted on a conventional constant acceleration drive. The spectra were fitted using the *SpectrRelax* program [5]. The isomer shift values are referred to that of $\alpha$-Fe at 300 K.

Density functional theory (DFT) calculations [6] of quadrupole splitting were used for study electronic state and crystal environment of iron in Fe$_3$PO$_7$. DFT calculations were made using the ORCA program [7]. Electric field gradient was calculated for the cluster, which contains 40 atoms using density B3LYP [8] functional available in the ORCA [7] package. The def2-TZVPP [9] basis for EFG tensors calculation was selected.

## 3 Results and discussions

The $^{57}$Fe Mössbauer spectra of Fe$_3$PO$_4$O$_3$ measured in the paramagnetic temperature range ($T > T_N$) (Fig. 2a) consist of a single quadrupole doublet with narrow ($W = 0.31(1)$ mm/s) and symmetrical lines, emphasizing the uniformity of structural positions of iron atoms in the phosphate [3]. The value of the isomer shift $\delta_{300K} = 0.33(1)$ mm/s corresponds to high-spin ions Fe$^{3+}$($d^5$, $S = 5/2$) located in oxygen (FeO$_n$) polyhedra with coordination number $n > 4$ [10]. It is interesting that the observed $\delta$ value is closer to typical values of isomer shifts ($\delta_{300K} \sim 0.36$ mm/s [10]) for Fe$^{3+}$ ions in octahedral oxygen surrounding where, in contrast to trigonal bipyramidal (FeO$_5$) polyhedra ($\delta_{300K} \sim 0.27$ mm/s [10]), the Fe-O bonds are considered to be almost entirely ionic. Apparently, this coincidence is due to the fact that the Fe←O transfer of the electron density in trigonal bipyramidal (FeO$_5$) polyhedra induces a simultaneous increase in the 4$s$ and 3$d$ orbital populations of iron ions, which affects the value of the isomer shift in the opposite directions [10, 11]. Mutual compensation of these two effects possibly renders the resulting isomer shift "less sensitive" to specific chemical bonds of iron in oxides, in contrast to quadrupole splitting which allows estimating the symmetry of local surrounding and spin state of Fe$^{3+}$ ions.

The high quadrupole splitting of the doublet $\Delta_{300K} = 1.10(1)$ mm/s shows that the $^{57}$Fe nuclei are located in crystal positions with a strong electric field gradient (EFG). As the EFG reflects the asphericity of the charge-density distribution near the probe $^{57}$Fe nucleus, it is directly related to the electron-density distribution and the symmetry in the nearest environment of the chemical bond. To get a better understanding of the charge redistribution effect in the Fe$_3$PO$_7$ lattice, we calculated the EFG using the density-functional theory (DFT) method [6]. The components ($V_{ij}$) of the EFG can be calculated directly from the electronic and nuclear charge distribution by [7]:

$$V_{ij} = -\sum_{kl} P_{kl} \langle \phi_k | (\frac{3r_i r_j - \delta_{ij} r^2}{r^5}) | \phi_l \rangle + \sum_{k(t)} Z_t (\frac{3\xi_{ki}\xi_{kj} - \delta_{ij} R_k^2}{R_k^5}), \quad (1)$$

where $r$ is the electronic position relative to the $^{57}$Fe nucleus, $R_k$ is the distance between $^{57}$Fe and nucleus "$t$", $r_{i(j)}$ and $\xi_{k,i(j)}$ ($i, j$ = x, y, z) are the projections of the vectors $\boldsymbol{r}$ and $\boldsymbol{R}_k$, respectively, $\phi_{k(l)}$ are the basis single-electron wave functions (for ex., "atomic" orbitals), $\{P_{kl}\}$ are the elements of the density matrice [7]. First contribution in (1) is the asymmetric distribution of the valence electrons ($V^{el}$) of the iron ion under consideration. Second, the lattice contribution ($V^{lat}$) arising from the point charges ($Z_t$) in the surroundings. The calculated $V^{el}$ and $V^{lat}$ values for the principal component $V_{ZZ}$ (where $|V_{ZZ}| \geq |V_{YY}| \geq |V_{XX}|$) are $V^{el}_{ZZ} = -9.40 \times 10^{21}$ V/m$^2$ and $V^{lat}_{ZZ} = -3.05 \times 10^{21}$ V/m$^2$. Obviously, the total $V_{ZZ}$ value is dominated by the valence electrons. To a good approximation, the valence contribution is determined by the anisotropy $\Delta n_p$ and $\Delta n_d$ of the $p$ and $d$ electrons [12]:

$$V^{el}_{ZZ} = a\Delta n_p + b\Delta n_d = a(\tfrac{1}{2}[n_{p_x} + n_{p_y}] - n_{p_z}) + b(n_{d_{x2-y2}} + n_{d_{xy}} - \tfrac{1}{2}[n_{d_{xz}} + n_{d_{yz}}] - n_{d_{z2}}), \quad (2)$$

where $a \propto \langle r^{-3}\rangle_p$ and $b \propto \langle r^{-3}\rangle_d$ are the average values of the inverse radial functions the $p$ and $d$ electrons. Using the Mulliken population analysis [13], we calculated the corresponding deviations $\Delta n_p = 0.018$ and $\Delta n_d = -0.035$ (Table S1). According to this evaluation, the anisotropy $\Delta n_d$ is much more pronounced than that of $\Delta n_p$, indicating stronger anisotropic spatial distribution of $d$ electrons compared with that of the $p$ electrons. The negative $V^{el}_{ZZ,3d} \propto \Delta n_d$ value shows that excess charges are accumulated in the $z$ direction and the contributions from $d_{xz}$, $d_{yz}$ and $d_{z2}$ orbitals dominate over those of $d_{x2-y2}$ and $d_{xy}$ orbitals oriented within the $xy$ plane. Such a redistribution of $d$ electrons qualitatively agrees with the local symmetry of the distorted trigonal bipyramid (FeO$_5$) clusters, where the apical Fe-O bonds are more elongated than equatorial ones. It should be noted, that although $\Delta n_d$ is significantly greater than $\Delta n_p$ the much larger expectation value $\langle 1/r^3\rangle_{np}$ of the $np$ radial functions causes the large contributions of the core $np$ electrons ($n$ = 1- 3) to the EFG.

Using the calculated EFG components $V_{XX}$ = 2.154×10$^{21}$ V/m$^2$, $V_{YY}$ = 4.638×10$^{21}$ V/m$^2$, $V_{ZZ}$ = -6.792×10$^{21}$ V/m$^2$ and the asymmetry parameter $\eta$ = 0.366 defined as the ratio ($V_{XX}$ – $V_{YY}$)/$V_{ZZ}$, we evaluated the quadrupole splitting $\Delta = eQV_{ZZ}/2(1 + \eta^2/3)^{1/2}$ (taking the value of $Q$ = +0.16 b for the nuclear quadrupole moment of the $^{57}$Fe [11]). A reasonable agreement between the experimental (1.01 mm/s) and theoretical (1.159 mm/s) values of $\Delta$ was obtained. The component $V_{ZZ}$ was found negative as well indicating that the quadrupole coupling constant $eQV_{ZZ}$ is negative (for the positive $Q$). The calculations revealed that the $V_{ZZ}$ component makes an angle of $\beta \approx 40^0$ with the $c$-axis in a hexagonal coordinate system (Fig. 2b) and a positive $V_{YY}$ component perpendicular to the $c$-axis, thus $V_{YY}$ is lying in the plane ($ab$) at about $60^0$ from the $b$ axis.

Below $T_N \approx 163$ K, a complex Zeeman magnetic structure appears in the Mössbauer spectra (Fig. 3). The observed inhomogeneous line broadenings of the spectra reflects a high degree of correlation between the values of the magnetic hyperfine field $H_{hf}$ at $^{57}$Fe nuclei and quadrupole shift ($\varepsilon_Q$) of the Zeeman components. It should be noted that a similar hyperfine magnetic structure was observed for the iron oxides BiFeO$_3$ [14, 15], AgFeO$_2$ [16] and FeVO$_4$ [17] possessing a non-collinear magnetic structure. To describe the inhomogeneous line broadening, we took into account the dependence of the quadrupole shift $\varepsilon_Q$ on the polar ($\theta$) and azimuthal ($\varphi$) angles of the magnetic hyperfine field $\boldsymbol{H}_{hf}$ with respect to the principal axes of the EFG tensor. The goodness of the fit has been significantly improved using the first ($\varepsilon_Q^{(1)}$) and second ($\varepsilon_Q^{(2)}$) order of perturbation theory [18]:

$$\varepsilon_Q^{(1)} = (-1)^{|m_I|+1/2}(\tfrac{1}{8}eQV_{ZZ})[3\cos^2\theta - 1 + \eta\sin^2\theta\cos 2\varphi]$$

$$\varepsilon_Q^{(2)} = (-1)^{|m_I|+1/2}\frac{(\tfrac{1}{4\sqrt{2}}eQV_{ZZ})^2}{g_{ex}\mu_n H_{hf}}\left([6 - 4\eta\cos 2\varphi]\cos^2\theta + \xi[\tfrac{3}{2}\sin^2\theta + \eta(1+\cos^2\theta)\cos 2\varphi]\right)\sin^2\theta,$$

(3)

where $m_I$ are magnetic quantum numbers; $eQV_{ZZ}$ is the quadrupole splitting constant, which equals that in the paramagnetic state ($T > T_N$) if there is no distortion of the crystal lattice at $T_N$, $\xi = \pm 1/2$ is a coefficient depending on the specific line in the Zeeman structure, $\mu_n$ is the nuclear Bohr magneton, and $g_{ex}$ is the gyromagnetic ratio of the excited state. In our model, we suggested also a spatial anisotropy of the hyperfine field $\mathbf{H}_{hf} \propto \tilde{A} \cdot \mathbf{S}$ at the $^{57}$Fe nuclei, which can be described as angular dependency of the hyperfine coupling tensor ($\tilde{A}$) in the system defined by principal axes of the EFG tensor (Fig. 2b). For systems where the anisotropy is not too large, in place of actual field $\mathbf{H}_{hf}$, to use the component $\tilde{A} \cdot \mathbf{S}$ parallel to the spin direction $h_\parallel \equiv H_{hf}(\|\mathbf{S})$. For general case, $A_{xx} \neq A_{yy} \neq A_{zz}$ the expression for $h_\parallel$ is written in the following form:

$$h_\parallel / S = \tilde{A}_{is} + 1/6 \tilde{A}_{an}(3\cos^2\theta - 1) + \eta_m \sin^2\theta \cdot \cos 2\varphi, \tag{4}$$

where $\tilde{A}_{is} = 1/3[A_{xx} + A_{yy} + A_{zz}]$ and $\tilde{A}_{an} = [2A_{zz} - A_{xx} - A_{yy}]$ are isotropic and anisotropic parts of the hyperfine coupling tensor $\tilde{A}$, respectively, $\eta_m = 1/2[A_{xx} - A_{yy}]$ is the magnetic asymmetry parameter. The both angles $\theta$ and $\varphi$ can be expressed by the angle $\vartheta$, describing position of magnetic moment vector $\mu_{Fe}$ on the rotation plane, and the Euler angles ($\alpha\beta\gamma$), describing relation between principal (*XYZ*) axes of the EFG tensor and the rotation plane (Fig. 2b). Therefore, the experimental spectrum is approximated as a superposition of the Zeeman patterns each of them is characterized by a different value of the rotation angle $\vartheta$, which varies continuously in $0 \leq \vartheta \leq 2\pi$ interval. The fitting was done using formulas (1-2) in different iron sites. Finally, to take into account anharmonicity (bunching) of the spatial distribution of the magnetic moments of Fe$^{3+}$, Jacobian elliptic function [16, 19] was used:

$$\cos\vartheta(x) = \text{sn}[(\pm 4K(m)/\lambda)x, m], \tag{5}$$

where $\lambda$ is the period of helicoid, $K(m)$ is the complete elliptic integral of the first kind, and $m$ is the anharmonicity parameter related to the distortion (anharmonicity) of the spiral structure [16].

According to the helicoidal magnetic structure of Fe$_3$PO$_7$ [2], the magnetic moments $\mu_{Fe}$ are constant in magnitude and rotate in the plane containing the hexagonal *c* axis. However, using only the neutron powder diffraction data, the determination of the helical plane direction (***n***) in the (*ab*) plane is difficult due to domain and powder averaging [2]. Therefore, applying the above fitting procedure, we systematically investigated a range of the angle $\beta$ ($50 \leq \beta \leq 90^0$), by taking two other Euler angles ($\alpha, \gamma$) as adjustable parameters. In additional to the usual variables ($\delta$, $eQV_{ZZ}$), assumed to be equal for all Zeeman subspectra, the principal components $\{A_{xx}, A_{yy}, A_{zz}\}$ of the tensor $\tilde{A}$ and anharmonicity parameter $m$ were used as adjustable parameters. The value of asymmetry parameter $\eta = 0.366$ evaluated from the point charge calculations was fixed during the fitting processes. The resulting $\chi^2$ values for the fits are showing in Fig. 4 as a function of the variation in the angle $\phi$ between the helical plane direction ***n*** and the projection of the of the principal component $V_{ZZ}$ on the (*ab*) plane (inset for Fig. 4). This angle is directly related to Euler angle $\phi = \cos^{-1}\{\cos\beta/0.643\}$ in accordance with directions of the EFG principal components calculated previously. It is clearly visible that the best fit is obtained when the helical plane direction ***n*** is oriented in the (*ab*) plane at $\phi = 55^0 \div 65^0$ (or about $30^0$ from the *a* axis). This fit to the 15 K spectrum, obtained in this way, is shown in Fig. 3. Then, we tentatively fitted the spectrum assuming that the ***n*** direction is perpendicular to the hexagonal (*ab*) plane that corresponds to the Euler angle $\beta \approx 50^0$ (Fig. 2b). However, in this case the fit was of much poorer quality ($\chi^2 \approx 5.64$) than previous ones, and this supposition should be ruled out. At this point we can note that the analysis of the complex Mössbauer spectra at $T \ll T_N$ not only confirms several features of the helicoidal magnetic structure proposed in [2] but also allows to refine the helical plane direction, which cannot be determined from neutron powder diffraction data.

The hyperfine parameters deduced from the spectrum at $T = 15$ K are $\delta = 0.43(1)$ mm/s, $eQV_{ZZ} = -2.28(2)$ mm/s, $W = 0.32(1)$ mm/s. The fitted quadrupole coupling constant $eQV_{ZZ}$ has the same order of magnitude as the double quadrupole splitting ($2\Delta_{170K} \approx 2.23$ mm/s) measured just above $T_N \approx 160$ K. Inequality of the principal components $A_{xx}S = 458(1)$ kOe, $A_{yy}S = 465(1)$ kOe and $A_{zz}S = 472.4(3)$ kOe reflect the spatial anisotropy of hyperfine field. Using these values, we traced a polar diagram of spatial anisotropy in the system defined by principal axes of the rhombic ($\eta \neq 0$) EFG tensor (Fig. 5a). The projections of the $H_{hf}$ vector on the helical plane generate an elliptical-like profile with the maximal $H_\alpha \approx 469.3(5)$ kOe and minimal $H_\beta \approx 472.1(5)$ kOe components (Fig. 5b). The diagonal components ($A_{ii}$) of the hyperfine tensor $\tilde{A}$ are composed of three contributions:

$$A_{ii} \approx P_{Fe}[-\frac{4\pi}{3}\rho(0) + \frac{1}{2}\langle r^{-3}\rangle_d (g_{ii}-2) + \sum_{\mu\nu} P^\sigma_{\mu\nu} \langle \phi_\mu | (\frac{3r_i^2 - r^2}{r^5}) | \phi_\nu \rangle], \qquad (6)$$

where $P_{Fe} = g_e g_N \beta_e \beta_N$ is the proportionality factor, $P^\sigma_{\mu\nu}$ is the spin-density matrix, $r$ is a radius-vector that points from the nucleus to the electron, $\rho(0)$ is the spin density at the iron nucleus, $\{g_{ii}\}_{i=x,y,z}$ are the components of the effective $\tilde{g}^{eff}$ tensor, $\{\phi_k, \phi_l, ...\}$ is the set of basis 3*d*-functions. The first term in Eq.6 is the isotropic Fermi contact contribution, which determines the sign of the hyperfine tensor.

The anisotropy arises from the second and the third terms in Eq. 6 corresponding to the orbital ($\tilde{A}^{orb}$) and the electronic dipolar ($\tilde{A}^{dip}$) components, respectively. The orbital term $\tilde{A}^{orb}$ corresponds to the magnetic field produced at the nucleus due to orbital currents. In general, the low-symmetry crystal field quenches the orbital angular moment, but the spin-orbit coupling restores it an amount of $L_i \propto (g_{ii} - 2)S$ [7]. Thus, the anisotropic orbital term arises from the anisotropy of the $\tilde{g}^{eff}$-tensor, which differs very little from ~ 2 for the high-spin $Fe^{3+}$ ions with $A_{1g}$ orbital singlet ground state. The dipolar term $\tilde{A}^{dip}$ does not vanish only when the spin density is aspherical. For isolated high-spin $Fe^{3+}$ ions having spherical $3d$ electron distribution this term turns into zero. However, covalent effects due to the anion-cation $Fe^{3+}$-$O^{2-}$ → $Fe^{2+}$-$O^{-}$(*L*) charge transfer in the low-symmetry distorted $FeO_n$ polyhedra produces inter-configurational mixing effects, in particular, mixing of the $(3d^5)^6A_{1g}$ term with the orbitally active $(d^6\underline{L})\ ^6T_{1g}/^6T_{2g}$ terms for the charge transfer configuration $d^6\underline{L}$, where $\underline{L}$ denotes the oxygen hole. As a result, we arrive at nonzero $\tilde{A}^{dip}$. To verify this assumption, we used the Mulliken populations ($n_{3d}^{\uparrow(\downarrow)}$) of the spin-polarised $3d$ orbitals (Table S1) obtained from the DFT calculations. Substituting these values into the expressions for the principal $A_{ii}^{dip}[n_{3d}^{\uparrow(\downarrow)}]$ components (Eq. S3), we evaluated the theoretical values $A_{xx}^{dip}S$ = -0.53 kOe, $A_{yy}^{dip}S$ = -0.04 kOe and $A_{zz}^{dip}S$ = 0.57 kOe, the difference between which is significantly less than the corresponding experimental values. The above estimates show that the observed in our experiments anisotropy of the hyperfine field at the $^{57}Fe$ nuclei is not related to the intrinsic features of iron electronic state in $Fe_3PO_7$ (non-quenched orbital contribution, reduced spin state, anisotropic charge-transfer ...) as was assumed previously in Ref. [2].

Another "external" source of an anisotropy of the local field $H_{hf}$ at the nucleus of the high-spin $Fe^{3+}$ ions is an dipole field $\tilde{A}^D$ induced by the neighboring magnetic ions:

$$A_{ii}^D = \frac{\mu_0}{4\pi} \sum_{k(t)} \frac{(3\xi_{ik}^2 - R_k^2)}{R_k^5} \cdot \mu_{ik}, \qquad (7)$$

where $k(t)$ is the summarized index on all positions of $t$-ion, $\xi_{ik} = \{x_k, y_k, z_k\}$ and $R_k$ are Descartes coordinates (refer to the principal EFG axes frame) and radius vector of $t$-ion in position $k(t)$, $\mu_{ik}$ is the projection of the iron magnetic moment. Substituted in the Eq. 4 the coordinates ($\xi_{ik}$) of the iron ions within the triangle plane and in different $c$-axis layers, taken from [2], the values of dipole contributions were evaluated: $A_{xx}^D$ = -2.1 kOe, $A_{yy}^D$ = -4.6 kOe and $A_{zz}^D$ = 6.7 kOe. Using these values, we calculated the anisotropic part of the hyperfine coupling tensor $A_{an}^D$ = 20 kOe associated with the dipole field, which is close to the experimental value of $\tilde{A}_{an}$ = 20.8(8) kOe from Eq. 4. We thus conclude that the main contribution to the observed spatial anisotropy of $H_{hf}$ is due to the anisotropy of the dipole field $\tilde{A}^D$ induced by the neighboring iron ions.

Since the large isotropic contribution to the internal $H_{hf}$ field resulting from the Fermi contact interaction of the intrinsic spin of the $Fe^{3+}$ ion with $ns$-electrons exceeds by two orders of magnitude the anisotropic contribution, one may take into account only the projections of the anisotropic fields ($h^D$) on the direction of the isotropic hyperfine field:

$$h^D(\vartheta) = \frac{\mu_0}{4\pi} \sum_{k(t)} \mu_k \frac{(3x_k'^2 - r_k^2)\cos^2\vartheta + (3y_k'^2 - r_k^2)\sin^2\vartheta + x_k' y_k' \cos\vartheta \sin\vartheta}{R_k^5}, \qquad (8)$$

where $\vartheta$ is the between the direction of $H_{hf}$ and the local $x'$ axis oriented perpendicular to the hexagonal $c$ axis, which is located within the $(x'y')$ plane of spin rotation (Fig. 5b), $\mu_0$ is the permeability constant. Using the above expression (in the coordinate system where iron ions spin rotates), we plotted a polar diagram $h^D(\vartheta)$ (in Fig. 5b), which qualitatively reproduces the distorted elliptic-like profile of the experimental hyperfine field $H_{hf}(\vartheta)$. Notice, the axes $H_\alpha$ and $H_\beta$ of the ellipse do not necessarily coincide with the principal axes $x'$ and $y'$ of the $\tilde{A}^{dip}$ tensor oriented in the $(x'y')$ plane at ~$20^0$ from the $c(\sim\|y')$ axis (Fig. 5b). Some deviations of these profiles may be related with the small contribution of the anisotropic field ~$(\tilde{A}^{orb} + \tilde{A}^{dip}) \cdot S$ due to the reduction of the symmetry of the $Fe^{3+}$ wave functions from cubic in the low-symmetry crystal field.

The use of the above model allowed to satisfactorily describe the entire series of experimental spectra measured in the magnetic ordering temperature range 15 K ≤ $T$ < $T_N$ (Fig. 3). We could not find any visible anomalies in the temperature dependences of hyperfine parameters (Fig. 6). The isomer shift $\delta(T)$ gradually decrease in accordance with the Debye approximation for the second-order Doppler shift [10] (Fig. 6a). The best fit for effective Debye temperature $\Theta_D$ = 552(3) K is in good agreement with the corresponding values for the $Fe^{3+}$ ions in other iron oxides [11]. In the same temperature range, the observed $eQV_{ZZ}(T)$ dependence (inset for Fig. 6a) is mainly due to the temperature variation of the lattice $V^{lat}(T)$ contribution to the EFG, which can be described using a semi-empirical relation $|eQV_{ZZ}(T)| = A(1 - B \cdot T^{3/2})$ with $A$ = 2.277(9) mm/s and $B$ = 5.8(8)$\cdot 10^{-6}$ $K^{-3/2}$.

Taking into account the pronounced temperature dependence of the $Fe_3PO_7$ crystal parameters near $T_N$ [2], the temperature dependence of magnetic field $H_{hf}(T) = S/3\{A_{xx}(T) + A_{yy}(T) + A_{zz}(T)\}$ can be analyzed using the Bean-Rodbell (B-R) model [20] than normal Brillouin function (Fig. 6b). In this approximation [20] the exchange magnetic interactions are considered to be a sufficiently strong function of the lattice spacing, and the hyperfine field $H_{hf}(T)$ is expressed as:

$$H_{hf}(T) = H_{hf}(0) \cdot B_S \left[ \frac{3S}{S+1} \frac{\sigma(T)}{\tau} \left( 1 + \frac{3}{5} \frac{(2S+1)^4 - 1}{2(S+1)^3 S} \zeta \sigma^2(T) \right) \right], \tag{9}$$

where $S$ = 5/2 is the total spin of the $Fe^{3+}$ ions, $\sigma(T)$ is reduced hyperfine field $H_{hf}(T)/H_{hf}(0)$, $\tau = T/T_N$ is the reduced temperature, and $H_{hf}(0)$ is the saturation hyperfine magnetic field, $\zeta$ is fitting parameter, which involves the magneto-structural coupling coefficient. The value of this parameter controls the order of the magnetic phase transition [20]. A reasonably good fit to the B-R model was obtained for the magnitude of $\zeta$ = 0.53(1) that indicates a second-order phase transition. We estimated the saturation field $H_{hf}(0)$ = 461.5(4) kOe and the point $T_N \approx$ 168(1) K, which is close to the Neel point (~163 K) found from magnetic measurements [20]. This shows that there are no any electronic and structural transitions in the whole magnetically ordered temperature range.

Alternatively, the temperature dependence of $H_{hf}(T)$ can be approximated using a power law $H_{hf}(T) = H_{hf}(0)(1 - (T/T_N)^\alpha)^\beta$, where $\beta$ is a critical exponent and $\alpha$ is an empirical fitting parameter to describe the experimental data well below $T_N$ ($T \ll T_N$). A reasonably good fit (Fig. 6 b) leads to $H(0) = 473(2)$ kOe, $T_N = 165.4(8)$ K, $\alpha = 1.5(1)$ and $\beta = 0.24(1)$. The value of the critical exponent well corresponds to the theoretical value $\beta^{th} = 0.23$ expected for a systems with a 2D *XY* magnets [21]. However, the origin of the critical parameter $\beta$ in the quasi-*2D* systems is far from trivial, due to competition between several interactions such as the magnetic coupling between layers and the strength of the crystal field (single-ion anisotropy), which can lead to a range of $\beta$ values in the range $0.20 \leq \beta \leq 0.36$. Further experimental and theoretical study is still required to reach a deeper understanding of critical dynamics and give a more definitive assignment of the universality class of this very interesting system.

In the temperature range near $T_N$ ($T \rightarrow T_N$) we observed the rapid broadening of the Zeeman lines and then appearance of a paramagnetic quadrupole doublet whose partial contribution sharply increases with temperature (Fig. 7). The hyperfine parameters ($\delta$, $\Delta$) of the doublet well correspond to those observed for the paramagnetic temperature range. Such a spectral behavior is characteristic for isolated superparamagnetic particles or nanosized magnetic domains with the randomly flipping direction of the magnetization under the influence of temperature. In the case of $Fe_3PO_7$ the needle-like domains can blocks of long-range magnetic order near the point $T_N$ suggesting strong thermal spin fluctuations. It should be noted, that the description of the experimental spectra was obtained by using anomalously high values of the anharmonicity parameter $m \approx 0.94(1)$, which remains almost constant in the range $T < T_N$. We can speculate that the strong anisotropy is also associated with the presence of the needle-like domains. The best fit of the spectra is obtained when the easy (bunching) axis is directed along the line of intersection of the helical plane and the hexagonal (*ab*) plane (Fig. 2b). It is possible that the domain walls create local stresses in the (*ab*) causing a non-uniform rotation of the iron spins.

Finally, we will comment on the lower value of the saturated magnetic field $H_{hf}(0) \approx 462$ kOe of $Fe_3PO_7$ in comparison to the 540-568 kOe values for the high-spin ferric ions in other 3D oxide systems [22]. This reduction can be related to the local magnetic surrounding of $Fe^{3+}$ ions via the transferred hyperfine field resulting from the all nearest ferric neighbors along the *c*-axis and within the (*ab*) plane. The experimental hyperfine field $\boldsymbol{H}_{hf}$ is the vector sum of two main contributions:

$$\boldsymbol{H}_{hf} = \boldsymbol{H}_{loc} + \sum_n B_n (\langle \boldsymbol{S}_n / S \rangle), \qquad (10)$$

where $\boldsymbol{H}_{loc} = \boldsymbol{H}_F + \boldsymbol{H}_{cov}$ is the local field that is the sum of the free-ion field $\boldsymbol{H}_F$, produced by the Fermi contact interaction, and the covalent contribution $\boldsymbol{H}_{cov}$, arising from the covalent transfer effects [23]. These two contributions are proportional to the vector <*S*> directed along the thermally averaged 3*d* spins. According to theoretical calculations [24, 25], $H_{loc} \approx (490 \div 500)$ kOe for $Fe^{3+}$ ions in octahedral coordination, and $H_{loc} \approx (410 \div 420)$ kOe for $Fe^{3+}$ ions located in tetrahedral oxygen surrounding. Unfortunately, we do not have any information concerning $Fe^{3+}$ in trigonal bipyramid. The second term in (10), is contribution resulting from all single-bridged nearest ferric neighbors "*n*", each proportional to the spin $<S_n>$, on the neighboring site, $B_n$ is a positive scalar parameter depending on the superexchange iron-oxygen-iron bond angle ($\psi$) and direct iron-iron bond distance [23]:

$$B_n = \left\{ (h_\sigma^{(n)} - h_\pi^{(n)}) \cos^2 \psi_n + h_\pi^{(n)} \right\}_{ST} + h_{dir}^{(n)}, \qquad (11)$$

where $h_\sigma$ and $h_\pi$ parameters arise from the supertransferred (ST) spin-polarization of iron *s*-orbitals, caused by the ligand *p*-orbitals that have been unpaired by spin transfer, via $\sigma$ and $\pi$ bonds, into unoccupied 3*d* orbitals on the neighboring cations; $h_{dir}$ is the direct contribution arising from the overlap distortions of iron *s*-orbitals by 3*d*-orbitals of the neighboring ions. The calculations of *Moskvin et al.* [25] for ferrites $RFeO_3$ have shown that $|h_\sigma| \approx 10$ kOe and $|h_\pi| \approx 1.6$ kOe. For the face-shared $FeO_5$ pyramids, due to the compensation effect of the weakened antiferromagnetic kinetic exchange with the ferromagnetic potential *s-d* exchange [25], the direct contribution $|h_{dir}|$ usually does not exceed 9-12 kOe [16].

$Fe_3PO_7$ exhibits noncollinear magnetic order, in which among six iron neighbors of the central iron there are two nearest neighbors with the same spin direction as the central ion and coupled with this one through direct exchange interactions (Fig. 1). The remaining four neighbors from triangular units in different *c*-axis layers, connected with the central ion by the supertransferred interactions Fe-O-Fe, have the opposite spin direction. Substituting in the Eq.11 the values of the angle $\psi \approx 125^0$ [15] and parameters $h_\sigma = 10$ kOe and $h_\pi = 2$ kOe [25], we evaluated the positive supertransferred contribution $H_{ST} = 4 \times B_{ST} \approx 18$ kOe from $Fe^{3+}$ ions in triangular units located in adjacent *c*-axis layers (a slight noncollinearity of neighboring iron moments leads to an error of not more than a few percent). This small positive contribution can be largely compensated by the negative "direct" contribution $H_{dir} = -2 \times h_{dir} \approx -(18 \div 24)$ kOe from the two neighbors with the same spin directions as the central iron ion, giving a total contribution of zero to the $H_{hf}$ field. As a result, the reduced $H_{hf}$ value for $Fe_3PO_7$ is presented as $H_{hf} \approx H_F + H_{cov}$, where the large negative contribution $H_{cov} = H_{hf} - H_F \approx -168$ kOe (we have chosen $H_F = 630$ kOe [26]) arises from the $Fe^{3+}$-$O^{2-} \rightarrow Fe^{2+}$-$O^{-}(\underline{L})$ charge transfer (where $\underline{L}$ denotes the oxygen hole). This conclusion agrees with the previously proposed explanation for the reduced effective moment per $Fe^{3+}$ ($\mu_{eff} \sim 4.2\ \mu_B$) [2] in $Fe_3PO_7$, attributed to the charge transfer in this insulating compound. Finally, the charge transfer $O^{2-} \rightarrow Fe^{3+}$ in the low-symmetry distorted $FeO_5$ polyhedra produces inter-configurational mixing effects, in particular, mixing of the $^6A_{1g}(d^5)$ term with the orbitally active ($<L> \neq 0$) $^6T_{1g}(d^6\underline{L})$ term, inducing the single-ion anisotropy for the charge transfer configuration $d^6\underline{L}$ [16].

We tried to fit the experimental Mössbauer spectra of $Fe_3PO_7$ assuming an alternative conical spin structure proposed in previous neutron diffraction study of polycrystalline samples $Fe_3PO_7$ [2]. This structure is characterized by the conical axis (***n***) directed along the crystal ***a*** axis (***n*** ∥ ***a***) and the opening angle $\alpha \approx 70^0$ (Fig. S2), which produces a good fit to the neutron diffraction pattern [2]. Fig. S.1 shows the fitted spectra (recorded at $T = 15$ K ) for different values of the anisotropic the hyperfine field $H_{an}$ (=$A_{an}S$) and the opening angle ($\alpha$) (a detailed description is given in *Supporting Information*). As we can see from this figure, despite the number of variable parameters this model does not allow to account for the shape of the experimental spectra. Therefore, the presented Mössbauer data allow us to give some preference to the helicoidal phase-modulated spin structure. A similar choice between the two above models is difficult to discern from the neutron powder diffraction data [2].

**Conclusions**

We have carried out detailed $^{57}Fe$ Mössbauer measurements on polycrystalline samples of $Fe_3PO_7$ that demonstrated the effectiveness of the suggested approach to analysis of the complex hyperfine magnetic structure of the spectra measured over a wide temperature range. The results presented above not only confirm several features of the helicoidal magnetic structure in $Fe_3PO_7$ but also allows refining the helical plane direction, which cannot be determined from neutron powder diffraction data. It has been shown that a good fitting of the experimental spectra can be achieved assuming that the electric hyperfine interactions are modulated when the $Fe^{3+}$ spin rotates with respect to the EFG axis and emergence of spatial anisotropy of the hyperfine field $H_{hf}$ at $^{57}Fe$ nuclei. The large anharmonicity parameter, $m \approx 0.94$, of the spiral spin structure resulting from easy-axis anisotropy in the plane of the iron spin rotation, can be related with the needle-like domains within the hexagonal (*ab*) plane. The Mössbauer spectra of $Fe_3PO_7$ cannot be described using alternative conical spin structure proposed in previous neutron diffraction studies. Analysis of the temperature dependence $H_{hf}(T)$ with the Bean-Rodbell model leads to the structural factor $\zeta \approx 0.53$ that suggests that magnetic phase transition is second-order in nature but with strong coupling magnetic ordering to the lattice deformation. The lower value of the saturated magnetic field $H_{hf}(0) \approx 462$ kOe is mainly related to the magnetic surrounding of $Fe^{3+}$ ions via the supertransferred hyperfine field and large negative contribution $H_{cov} \approx -168$ kOe arising from the $Fe^{3+}-O^{2-} \rightarrow Fe^{2+}-O^-(\underline{L})$ charge transfer. The DFT calculations yield reliable charge distribution to which the EFG is so sensitive. It was shown that, in addition to lattice contribution $V^{lat}$, very large weight has electronic contribution $V^{el}$, arising from the asymmetric distribution of the *p* core and 3*d* valence electrons. From the calculated occupation number, the fluctuation of $\Delta n_d$ is more pronounced than that of $\Delta n_p$, indicating stronger anisotropic spatial distribution of $Fe^{3+}$ $3d$ electrons. It was found that $3d$ electrons move from $d_{x2-y2}$ and $d_{xy}$ orbitals to $d_{xz}$, $d_{yz}$ and $d_{z2}$ orbitals, which correlates with the local symmetry of the distorted trigonal bipyramid $(FeO_5)$ clusters. Such electronic redistribution in the low-symmetry crystal field may give the small anisotropic contribution $\sim(\tilde{A}^{orb} + \tilde{A}^{dip}) \cdot S$ in the hyperfine field $H_{hf}$ at $^{57}Fe$ nuclei. However, we have shown that the main contribution to the observed anisotropy of $H_{hf}$ is due to the anisotropy of the dipole field $\tilde{A}^D$ induced by the neighboring iron ions.

**Supporting Information Available:** Figures S1, S2, and Table S1, additional details of Mössbauer simulation using conical spin structure, and additional details and results from DFT calculations.

## FIGURES CAPTIONS

**Figure 1.** Schematic magnetic surrounding of $Fe^{3+}$ in $Fe_3PO_7$ (exchange interactions and integrals within ($J_1$) and between ($J_2$) the triangles $(Fe)_3$ along the $c$-axis are shown). Schematic representation of different contributions ($h_{ST}$ and $h_{dir}$) to the $H_{hf}$ value for the phosphate.

**Figure 2.** (a) $^{57}Fe$ Mössbauer spectrum of $Fe_3PO_7$ recorded at $T = 298$ K ($T >> T_N$) (the solid red line is the result of simulation of the experimental spectra). (b) Schematic view of the local crystal structure of $Fe_3PO_7$ (in hexagonal base) and directions of the principal EFG $\{V_{ii}\}_{i=x,y,z}$ axes ($\{\alpha,\beta,\gamma\}$ are the Euler angles, describing relation between principal ($XYZ$) axes of the EFG tensor and the rotation plane).

**Figure 3.** $^{57}Fe$ Mössbauer spectra (experimental hollow dots) of $Fe_3PO_7$ recorded at the indicated temperatures below $T_N$. Solid red lines are simulation of the experimental spectra as described in the text.

**Figure 4.** The resulting $\chi^2$ values for the fits of the spectra as a function of the variation in the angle ($\phi$) between the helical plane direction $n$ and the projection of the principal component $V_{ZZ}$ on the ($ab$) plane (see the inset). The $\phi$ angle is directly related to Euler angle $\phi = \cos^{-1}\{\cos\beta/0.643\}$.

**Figure 5.** (a) Surface plot of the $\tilde{A}$ tensor relative to the principal EFG $\{V_{ii}\}_{i=x,y,z}$ axes, and (b) the elliptic-like contour of this function in the rotation plane of magnetic moments of iron ions. Distribution of $H_{hf} \propto \tilde{A} \cdot S$ was calculated using formula (4) in 250 different iron sites (represented schematically by blue arrows) and taking the anharmonicity parameter $m \approx 0.94$ ($H_\alpha$ and $H_\beta$ denote the maximal ($H_\alpha$) and minimal ($H_\beta$) anisotropic hyperfine field components). The red dashed line corresponds to the projection of the dipolar field $h^D(\vartheta)$ (*see* Eq. 8) on the spin rotation plane.

**Figure 6.** (a) Isomer shift $\delta(T)$ as a function of the reduced temperature (red solid line corresponds to the Debye approximation for the second-order Doppler shift), and quadrupole coupling constant $eQV_{ZZ}$ plotted versus temperature (red solid line corresponds to the fitting using semi-empirical relation, *see text*). (b) Temperature dependence of hyperfine magnetic field $H_{hf}(T)$. The dark broken line shows the Brillouin function with $S = 5/2$; the blue solid line corresponds to fit using the Bean-Rodbell model; dashed dot line indicates the power law fit model.

**Figure 7**. $^{57}$Fe Mössbauer spectra (experimental hollow dots) of $Fe_3PO_7$ recorded near the Neel temperature ($T \to T_N$). Solid lines are simulation of the experimental spectra as the superposition of magnetic (red line) and paramagnetic (blue line) subspectra.

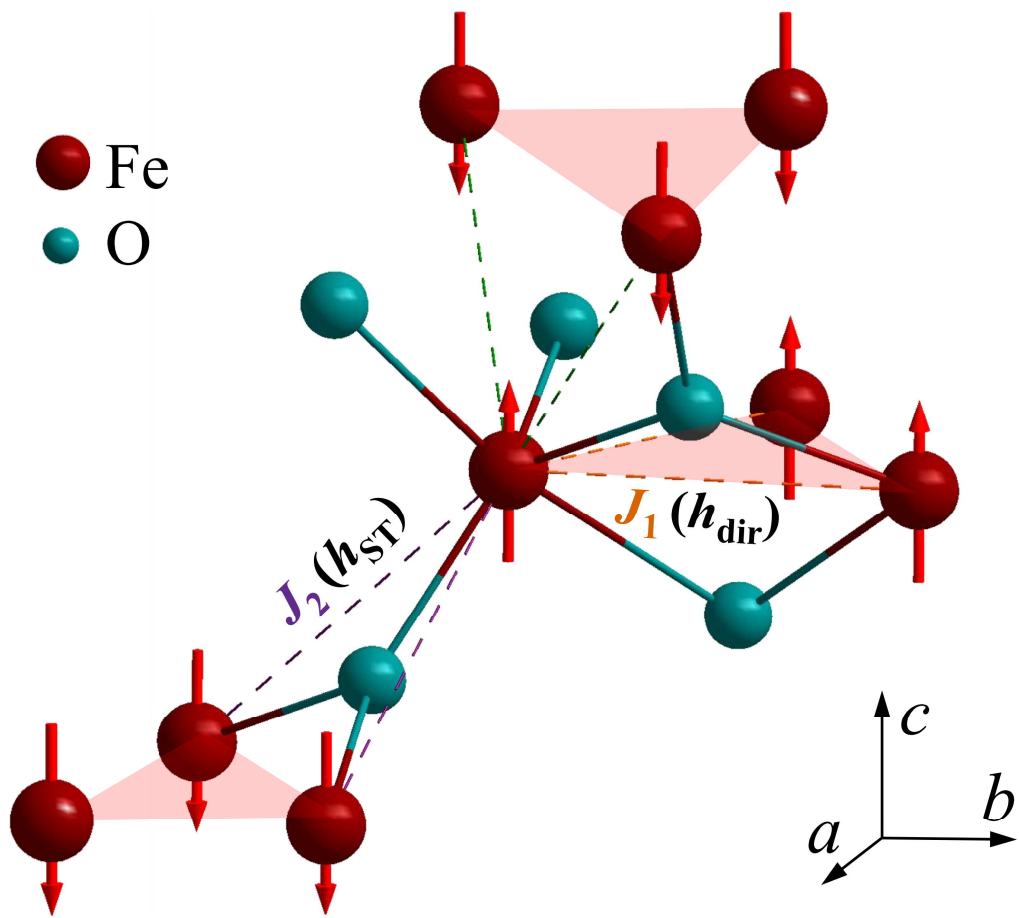

**Figure 1.** Schematic magnetic surrounding of $Fe^{3+}$ in $Fe_3PO_7$ (exchange interactions and integrals within ($J_1$) and between ($J_2$) the triangles $(Fe)_3$ along the $c$-axis are shown). Schematic representation of different contributions ($h_{ST}$ and $h_{dir}$) to the $H_{hf}$ value for the phosphate.

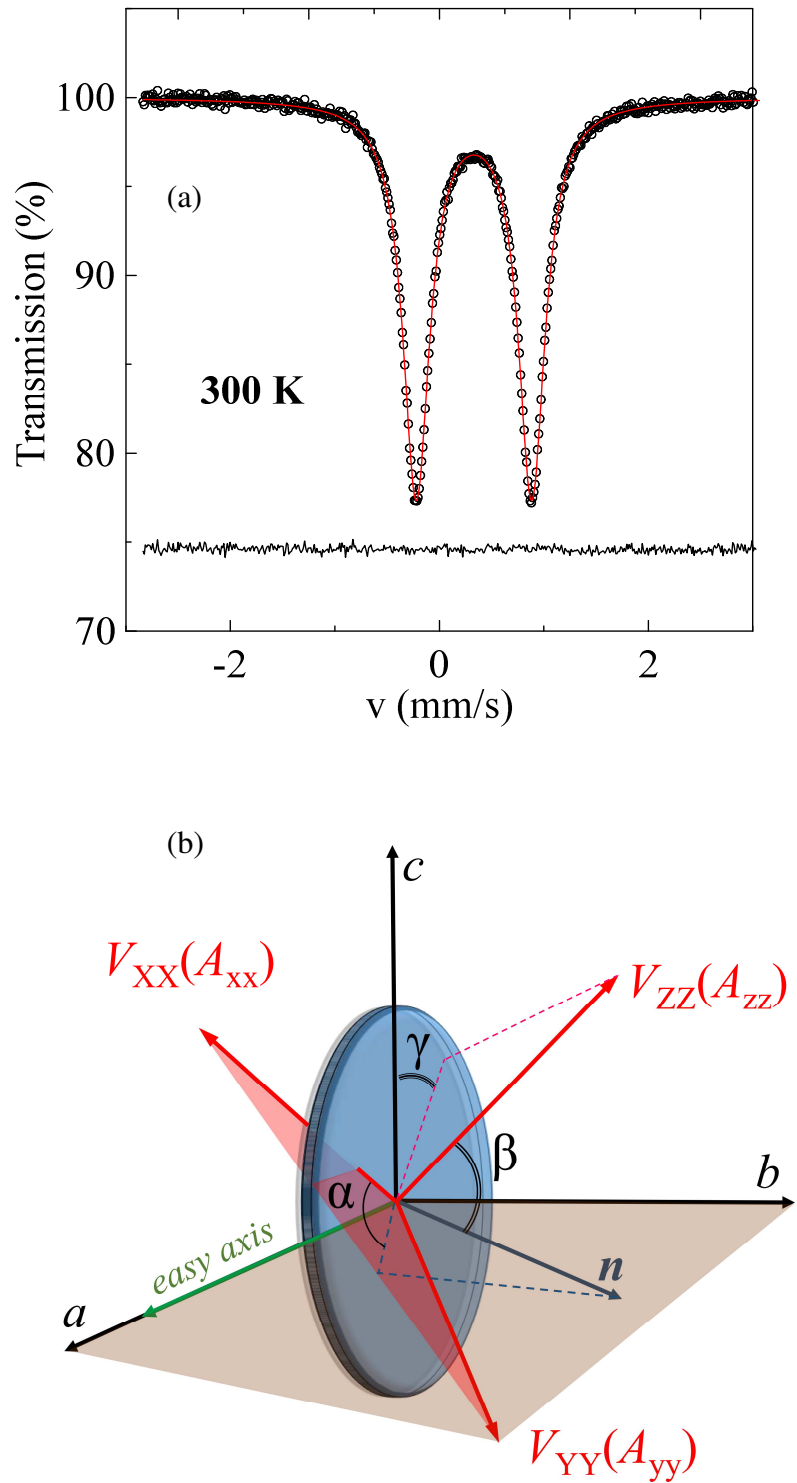

**Figure 2.** (a) $^{57}$Fe Mössbauer spectrum of $Fe_3PO_7$ recorded at $T = 298$ K ($T \gg T_N$) (the solid red line is the result of simulation of the experimental spectra). (b) Schematic view of the local crystal structure of $Fe_3PO_7$ (in hexagonal base) and directions of the principal EFG $\{V_{ii}\}_{i=x,y,z}$ axes ($\{\alpha,\beta,\gamma\}$ are the Euler angles, describing relation between principal (XYZ) axes of the EFG tensor and the rotation plane).

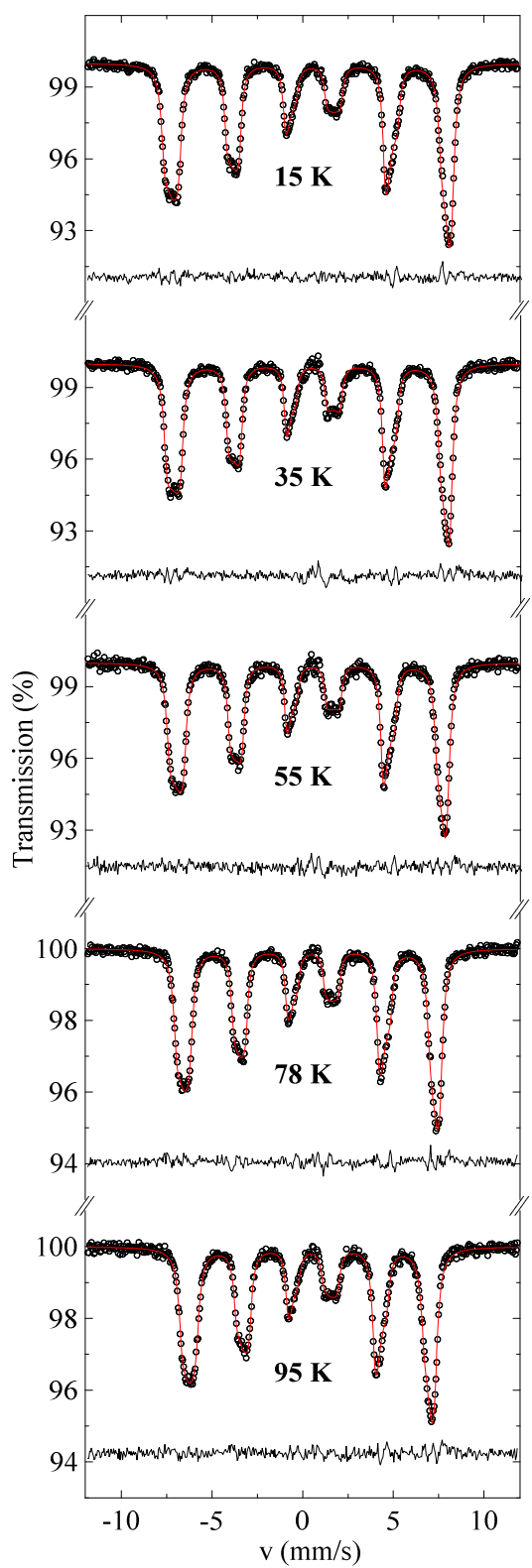

**Figure 3**. $^{57}$Fe Mössbauer spectra (experimental hollow dots) of Fe$_3$PO$_7$ recorded at the indicated temperatures below $T_N$. Solid red lines are simulation of the experimental spectra as described in the text.

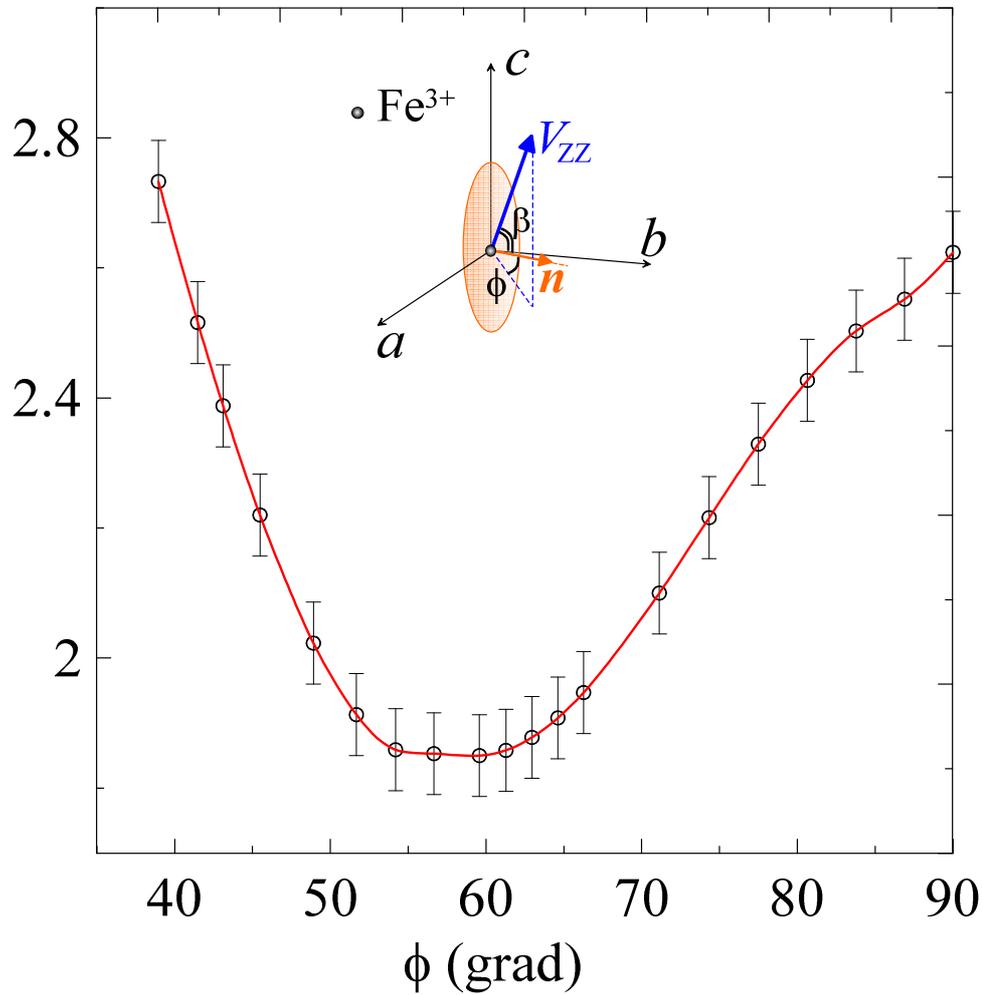

**Figure 4.** The resulting $\chi^2$ values for the fits of the spectra as a function of the variation in the angle ($\phi$) between the helical plane direction ***n*** and the projection of the of the principal component $V_{ZZ}$ on the (*ab*) plane (see the inset). The $\phi$ angle is directly related to Euler angle $\phi = \cos^{-1}\{\cos\beta/0.643\}$.

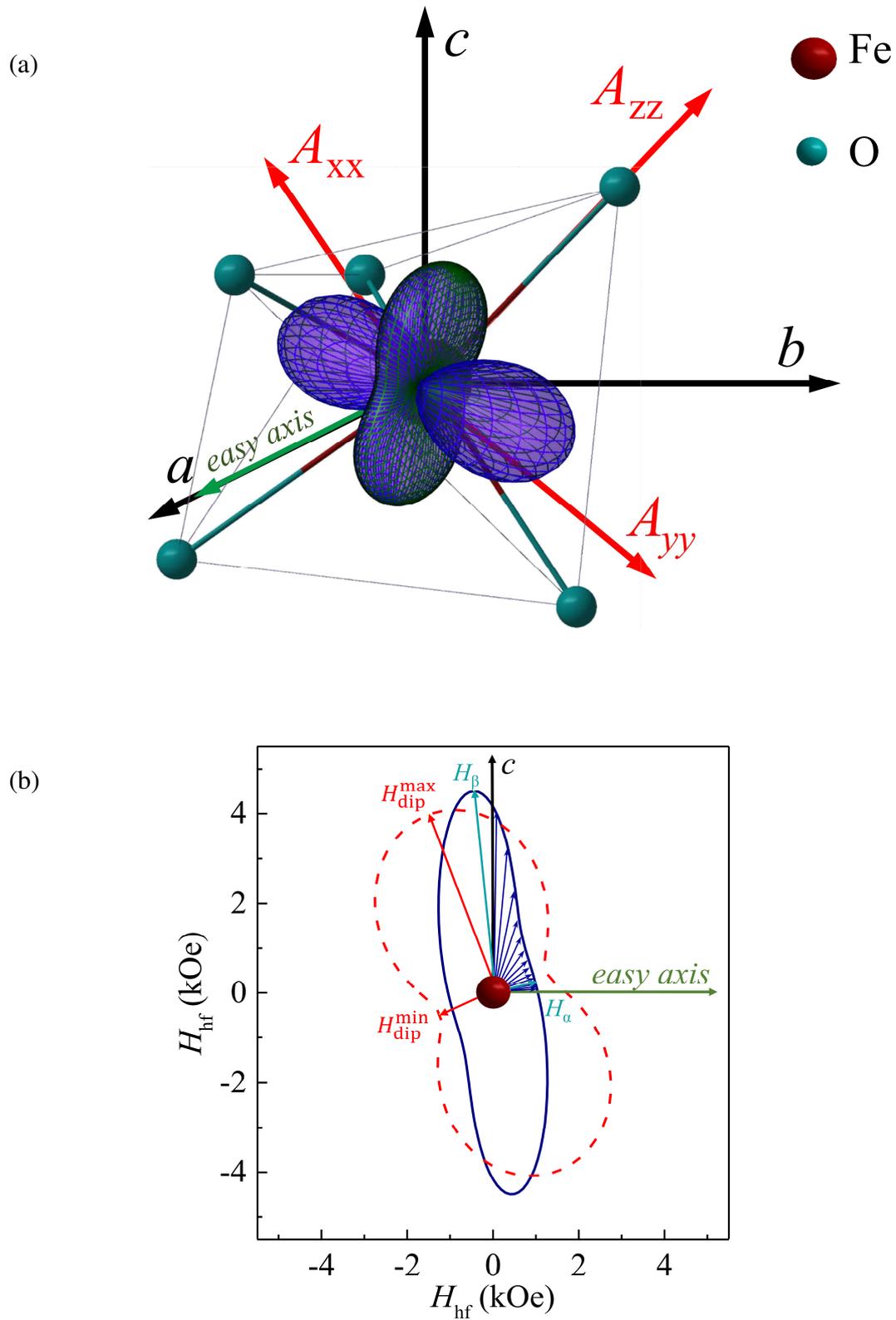

**Figure 5.** (a) Surface plot of the $\tilde{A}$ tensor relative to the principal EFG $\{V_{ii}\}_{i=x,y,z}$ axes, and (b) the elliptic-like contour of this function in the rotation plane of magnetic moments of iron ions. Distribution of $H_{hf} \propto \tilde{A} \cdot S$ was calculated using formula (4) in 250 different iron sites (represented schematically by blue arrows) and taking the anharmonicity parameter $m \approx 0.94$ ($H_\alpha$ and $H_\beta$ denote the maximal ($H_\alpha$) and minimal ($H_\beta$) anisotropic hyperfine field components). The red dashed line corresponds to the projection of the dipolar field $h^D(\vartheta)$ (*see* Eq. 8) on the spin rotation plane.

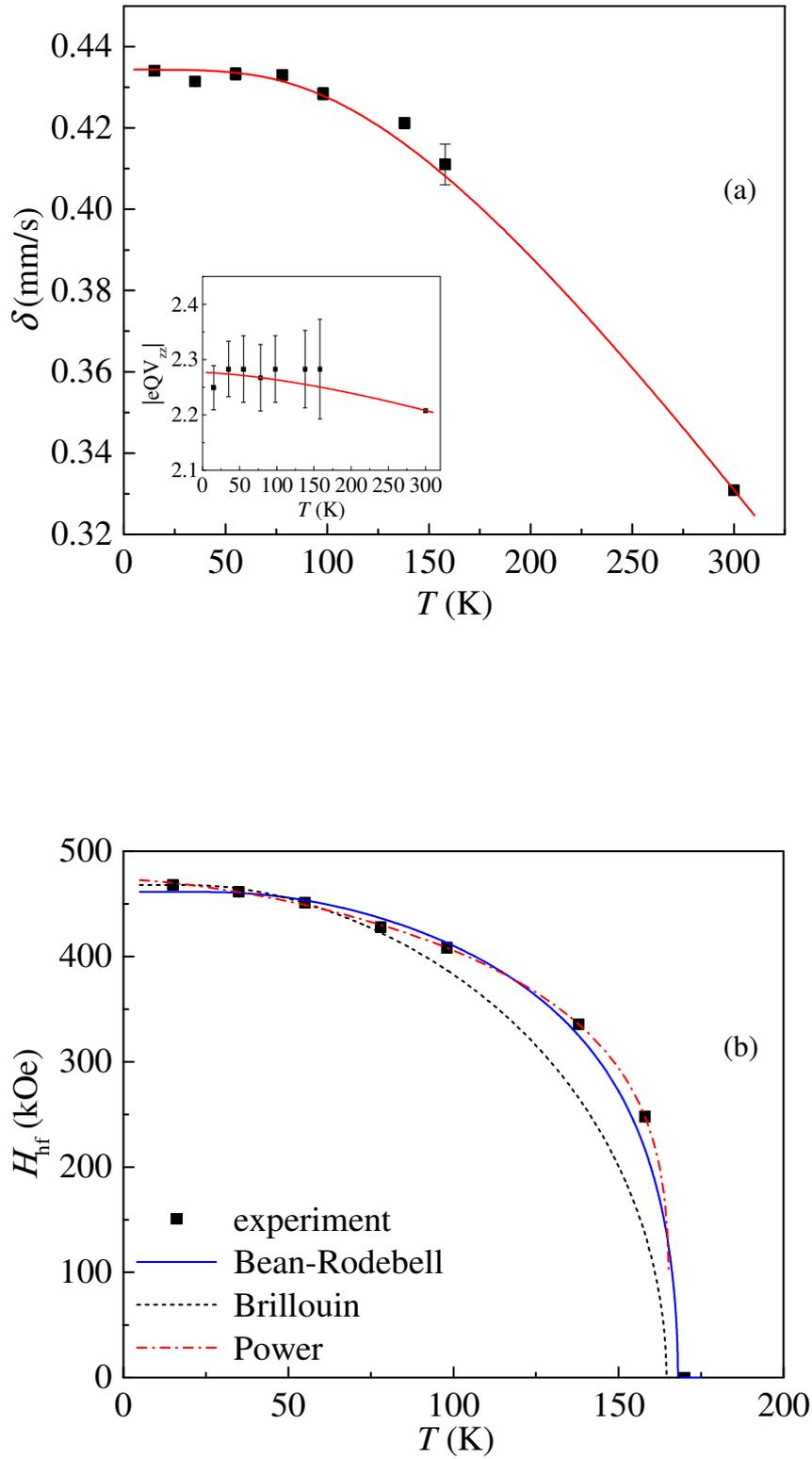

**Figure 6.** (a) Isomer shift $\delta(T)$ as a function of the reduced temperature (red solid line corresponds to the Debye approximation for the second-order Doppler shift), and quadrupole coupling constant $eQV_{ZZ}$ plotted versus temperature (red solid line corresponds to the fitting using semi-empirical relation, *see text*). (b) Temperature dependence of hyperfine magnetic field $H_{hf}(T)$. The dark broken line shows the Brillouin function with $S = 5/2$; the blue solid line corresponds to fit using the Bean-Rodbell model; dashed dot line indicates the power law fit model.

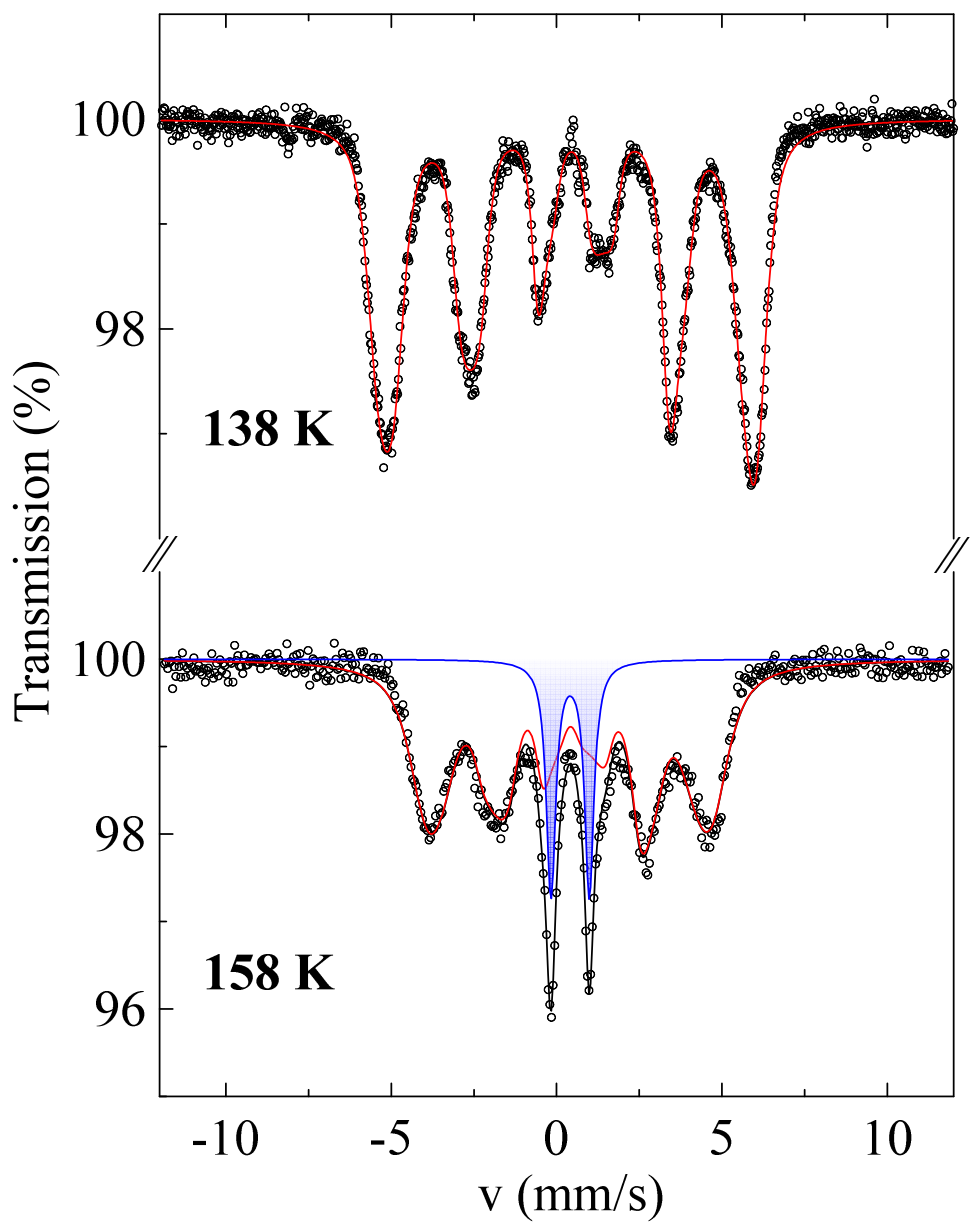

**Figure 7**. $^{57}$Fe Mössbauer spectra (experimental hollow dots) of $Fe_3PO_7$ recorded near the Neel temperature ($T \to T_N$). Solid lines are simulation of the experimental spectra as the superposition of magnetic (red line) and paramagnetic (blue line) subspectra.

# Modulated magnetic structure of Fe$_3$PO$_7$ as seen by $^{57}$Fe Mössbauer spectroscopy

A.V. Sobolev, A.A. Akulenko, I.S. Glazkova, D.A. Pankratov, I.A. Presniakov

*M.V. Lomonosov Moscow State University, Moscow, Russia*

*Supporting Information*

## 1. Mulliken orbital populations

**Table S1.** Mulliken μ-orbital populations $n_\mu^\sigma = P_{\mu\mu}^\sigma$, where the density matrices ($P_{\mu\nu}^\sigma$) are defined as: $P_{\mu\nu}^\sigma = \sum_{i=1}^{N_\sigma} C_{\mu i}^\sigma C_{\nu i}^\sigma$, where $N_\sigma$ is the number of electrons with spin σ ("↑" or "↓"), the index "i" runs over all molecular orbitals (MO). The contribution of the basis atomic $\phi_\mu$ orbital to the i$^{th}$ MO is equal to the square of the corresponding coefficients $(C_{\mu i}^\sigma)^2$.

| Orbital populations ($n_\mu^\sigma$) | 3*d*-orbitals | | | | | (2, 3)*p*-orbitals | | |
|---|---|---|---|---|---|---|---|---|
| | $d_{x2-y2}$ | $d_{z2}$ | $d_{xz}$ | $d_{yz}$ | $d_{xy}$ | $p_x$ | $p_y$ | $p_z$ |
| $n_\mu^\uparrow + n_\mu^\downarrow$ | 1.1487 | 1.1506 | 1.2163 | 1.1546 | 1.1520 | 4.0487 | 4.0229 | 4.0178 |
| $n_\mu^\uparrow - n_\mu^\downarrow$ | 0.8543 | 0.8529 | 0.8022 | 0.8542 | 0.8543 | 0.0104 | 0.0206 | 0.0207 |

## 2. Magnetic Hyperfine Interactions (spin-dipolar interaction)

The local spin dipole part arises from the magnetic dipole interaction of the magnetic nucleus with the magnetic moment of the electron (Eq. 6). The ligand field theory provides the folowing expressions for the dipolar contribution that can be used for the qualitative discussion (*see text*):

$$P_{Fe}[\sum_{kl} P_{kl}^\sigma \langle \phi_k | (\frac{3r_i^2 - r^2}{r^5}) | \phi_l \rangle] \Rightarrow \begin{cases} A_{xx}^{dip} = P_{Fe}\langle r^{-3}\rangle_d [\frac{2}{7}(n_{xz}^\sigma + n_{x2-y2}^\sigma + n_{xy}^\sigma - n_{z2}^\sigma) - \frac{4}{7}n_{yz}^\sigma] \\ A_{yy}^{dip} = P_{Fe}\langle r^{-3}\rangle_d [\frac{2}{7}(n_{yz}^\sigma + n_{x2-y2}^\sigma + n_{xy}^\sigma - n_{z2}^\sigma) - \frac{4}{7}n_{xz}^\sigma] \\ A_{zz}^{dip} = P_{Fe}\langle r^{-3}\rangle_d [\frac{2}{7}(n_{xz}^\sigma + n_{yz}^\sigma) + \frac{4}{7}(n_{z2}^\sigma - n_{x2-y2}^\sigma - n_{xy}^\sigma)] \end{cases} \quad (S1)$$

where $P_{Fe} = g_e g_N \beta_e \beta_N \approx 17.25$ MHz/au$^{-3}$, $\langle r^{-3}\rangle_d$ is the average value of the inverse cubic of the distance between 3*d* electron and the nucleus ($\langle r^{-3}\rangle_{d,(Fe^{3+})} \approx 5.63 - 6.68$ a.u.$^{-3}$ [11]).

## 3. Mössbauer fitting using the conical spin structure

We performed least-square-fitting of the Mössbauer spectrum recorded at 13 K assuming an alternative conical magnetic spin structure in Fe$_3$PO$_7$ [2].

Assuming that a hyperfine coupling tensor, $\tilde{A}$, specifying the coupling between the nuclear spin and the electronic spin ($S$), is diagonal with respect to the principal axes of the EFG tensor with axial symmetry, the hyperfine field $\boldsymbol{H}_{hf}$ can be written as follows,

$$\boldsymbol{H}_{hf} = \tilde{A}\cdot \boldsymbol{S} = iA_{XX}S_X + jA_{YY}S_Y + kA_{ZZ}S_Z, \qquad (S2)$$

where $S_X = S\cdot\sin\theta\cos\varphi$, $S_Y = S\cdot\sin\theta\sin\varphi$ and $S_Z = S\cdot\cos\theta$ are projections of Fe spin vector $\boldsymbol{S}$ on the principal axes of the EFG tensor ($\theta$ and $\varphi$ are the polar and azimuthal angles in the expressions for the first $\varepsilon_Q^{(1)}$ and second $\varepsilon_Q^{(2)}$ order quadrupole shift of the Zeeman components (*see text*, Eq. 3)), $A_{ii}$ are the values of the principal components of the hyperfine coupling tensor $\tilde{A}$. The expression (S2) suggests that the observed anisotropy of the hyperfine field $\boldsymbol{H}_{hf}$ can be analyzed in terms of anisotropy ($A_{XX} \neq A_{YY} \neq A_{ZZ}$) of the hyperfine coupling $\tilde{A}$ tensor. In this formalism, the value of the hyperfine field $\boldsymbol{H}_{hf}(\theta, \varphi)$ can be represented as

$$H_{hf} = \sqrt{A_{XX}^2 S_X^2 + A_{YY}^2 S_Y^2 + A_{ZZ}^2 S_Z^2} = \sqrt{A_{XX}^2 (\sum_i C_{ix'}S_{x'})^2 + A_{YY}^2 (\sum_i C_{iy'}S_{y'})^2 + A_{ZZ}^2 (\sum_i C_{iz'}S_{z'})^2}, \quad (S3)$$

where $S_{x'} = S\cdot\sin\alpha\cos\vartheta$, $S_{y'} = S\cdot\sin\alpha\sin\vartheta$, and $S_{z'} = S\cdot\cos\alpha$ are the projections of the spin vector $\boldsymbol{S}$ in the cylindrical coordinate system $x'y'z'$ for cone arrangement of magnetic moments, which tip out from cone symmetry axis ($N$) to an angle $\alpha$ and rotate about this axis by an angle $\vartheta$ (Fig. S1). The elements $\{C_{ij}\}$ of the $(x'y'z') \Rightarrow (XYZ)$ transformation matrix $\boldsymbol{C}$ are expressed in terms of three Euler angles $\{\Theta, \Phi, \psi\}$. In our case, the angle $\psi$ has a fixed zero value and the following two angles serve as the adjustable parameters: $\Theta$ is the azimuth angle between the directions of the principal component $V_{ZZ}$ and the $N$ axis, and $\Phi$ is the angle that determines the orientation of the projection of the $N$ axis onto the $(XY)$ plane (Fig. S1).

Using the above transformation, we found the expressions $\theta_{\Theta\Phi\alpha}(\vartheta)$ and $\varphi_{\Theta\Phi\alpha}(\vartheta)$ connected the spherical angles in the quadrupole shifts $\varepsilon_Q^{(1)}$ and $\varepsilon_Q^{(2)}$ (Eq. 3) with the rotation angle ($\vartheta$) of the iron magnetic moments.

$$\begin{pmatrix}\sin\theta\cos\varphi \\ \sin\theta\sin\varphi \\ \cos\theta\end{pmatrix} = \begin{pmatrix}\cos\Theta\cos\Phi & -\cos\Theta\sin\Phi & \sin\Theta \\ \sin\Phi & \cos\Phi & 0 \\ -\sin\Theta\cos\Phi & \sin\Theta\sin\Phi & \cos\Theta\end{pmatrix}\begin{pmatrix}\sin\alpha\cos\vartheta \\ \sin\alpha\sin\vartheta \\ \cos\alpha\end{pmatrix}, \qquad (S3)$$

where the ($\Theta, \Phi, \alpha$) angles serve as the fitted parameters that determine the orientation of the cone ($\Theta, \Phi$) in the EFG coordinate system $XYZ$, as well as the parameters of the cone itself ($\alpha$). The values $\vartheta$ is varying continuously between 0 and $2\pi$ depending on the coordinates of iron ions along the helix propagation (the rotation step $\Delta\vartheta$ was assume to be $\sim\pi/40$ that is sufficient to reproduce for such resolved Mössbauer spectra). The final expressions $\varepsilon_Q(\vartheta_{\Theta\Phi\alpha})$ and $H_{hf}(\vartheta_{\Theta\Phi\alpha})$ used in analyzing the spectra were obtained by substituting the dependencies $\theta_{\Theta\Phi\alpha}(\vartheta)$ and $\varphi_{\Theta\Phi\alpha}(\vartheta)$ into equations (3) and (S3), respectively.

The procedure of spectrum fitting was performed in three ways as it is shown at Fig. S1 supposing the direction of cone axis parallel to *a* crystal axis: (a) $\alpha$ angle is set as $70^0$ as it was proposed in [2] and the hyperfine field anisotropy was no taken into account; (b) $\alpha$ angle is set as $70^0$ and the hyperfine field anisotropy was taken into account in accordance with Eq. 4 (*see text*); (c) angle is not fixed and the hyperfine field anisotropy was taken into account in accordance with Eq. 4. One can noticed that there is no sufficient fit quality in all three approaches (Fig. S2) thus excluding the cone model of the magnetic structure of $Fe_3PO_7$.

**Figure S1.** Schematic view of the cylindrical coordinate system ($x'y'z'$) and directions of the principal EFG $\{V_{ii}\}_{i=X,Y,Z}$ axes ($\{\Theta, \Phi, \psi = 0\}$ are the Euler angles, describing relation between principal (*XYZ*) axes of the EFG tensor and the rotation plane).

**Figure S2**. $^{57}$Fe Mössbauer spectra (experimental hollow dots) of $Fe_3PO_7$ recorded at 13 K. Solid red lines are simulation of the experimental spectra assuming a conical magnetic spin structure in $Fe_3PO_7$ as described in the text.

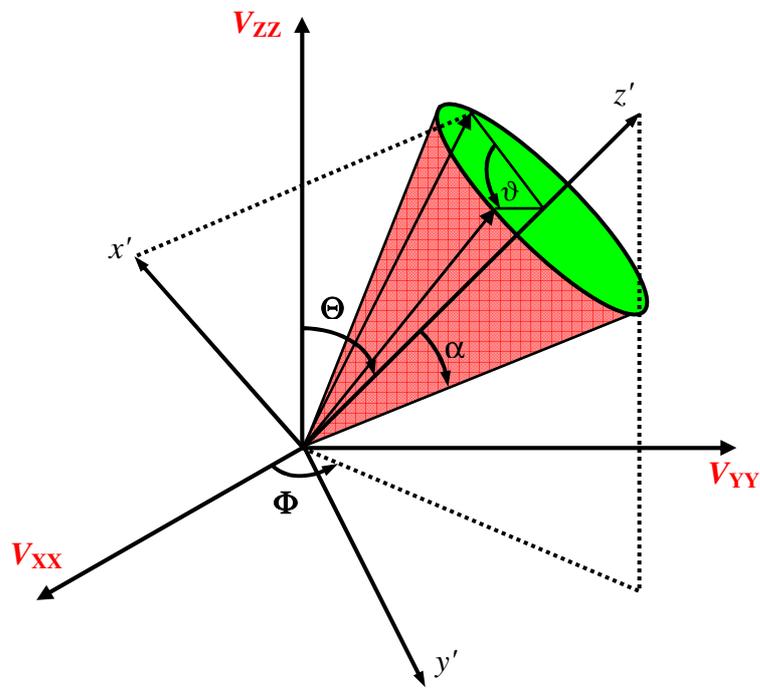

**Figure S1.** Schematic view of the cylindrical coordinate system ($x'y'z'$) and directions of the principal EFG $\{V_{ii}\}_{i=X,Y,Z}$ axes ($\{\Theta, \Phi, \psi = 0\}$ are the Euler angles, describing relation between principal (*XYZ*) axes of the EFG tensor and the rotation plane).

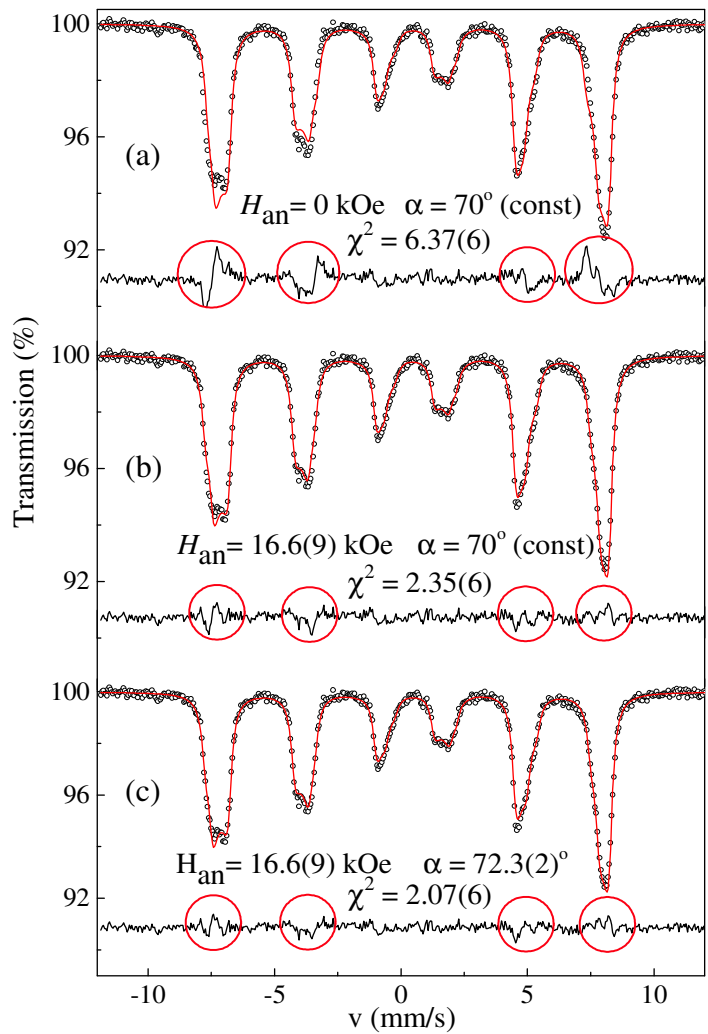

**Figure S2.** $^{57}$Fe Mössbauer spectra (experimental hollow dots) of Fe$_3$PO$_7$ recorded at 15 K. Solid red lines are simulation of the experimental spectra assuming a conical magnetic spin structure in Fe$_3$PO$_7$ as described in the text. $H_{an}$ is anisotropic part of the hyperfine field as described in eq. (4) ($H_{an} = S[2A_{zz} - A_{xx} - A_{yy}]$).